\documentclass[10pt,journal]{IEEEtran}

\usepackage[latin1]{inputenc}
\usepackage[english]{babel}
\usepackage[cmex10]{amsmath}
\usepackage{amssymb}
\usepackage{algorithm}
\usepackage{algpseudocode}
\usepackage{rotating}
\usepackage{enumitem}
\usepackage{MnSymbol}
\usepackage[usenames,dvipsnames]{xcolor}
\usepackage{fancybox}
\usepackage{array}
\usepackage{adjustbox}
\usepackage{cite}

\newcommand{\Z}{\mathbb{Z}}

\newcommand{\alg}[1]{\mathsf{#1}}
\newcommand{\sch}[1]{\mathsf{#1}}

\newcommand{\rv}[1]{{#1}}

\newtheorem{definition}{Definition}[section]
\newtheorem{proposition}{Proposition}[section]

\begin{document}

\title{Symmetric Blind Decryption with Perfect Secrecy}

\author{Juha~Partala\thanks{
J.~Partala is with the Department of Computer Science and Engineering, University of Oulu, Finland (e-mail: juha.partala@ee.oulu.fi).}}

\maketitle

\begin{abstract}
A blind decryption scheme 
enables a user to query decryptions from a decryption server without revealing
information about the plaintext message.
Such schemes are useful, for example, for the implementation of privacy preserving
encrypted file storages and
payment systems. 
In terms of functionality,
blind decryption is close to oblivious transfer.
For noiseless channels, information-theoretically secure oblivious transfer is impossible.
However, in this paper we show that this is not the case for blind decryption.
We formulate a definition of perfect secrecy of symmetric blind decryption for the following setting:
at most one of the scheme participants is a malicious observer.
We also devise a symmetric blind decryption scheme
based on modular arithmetic on a ring $\Z_{p^2}$, where $p$ is a prime,
and show that it satisfies our notion of perfect secrecy.
\end{abstract}

\begin{IEEEkeywords}
Communication system security, Cryptography, Encryption, Information security
\end{IEEEkeywords}


%

\section{Introduction}

%
Over the past 15 years, data has moved from local storage to
centralized data warehouses in the cloud.
The 
accessibility
of large amounts of personal data through a public network has given rise to
many security and privacy issues~\cite{Thuraisingham_2015}.
Fortunately, such issues have generally been taken seriously.
For example,
ethical and legal requirements have been imposed on guaranteeing
the confidentiality of medical records~\cite{HIPAA,DataProtDirective}.
However, the implementation of privacy technologies is non-trivial,
especially if the data storage has been outsourced to a cloud operator.
Sensitive information can often be inferred from simple access patterns either
by outsiders or by the operator of the storage.
For example, being able to observe a medical doctor
to access the medical record of a patient can leak sensitive information.
Therefore, such access patterns should be kept hidden both from
outsiders and from the party that is administering the records.

Oblivious databases~\cite{Coull_2009} and privacy-preserving encrypted filesystems~\cite{Green_2011}
are examples of technologies that can be used 
to hide the access information from the administrator.
For such systems, the decryption of data is typically handled by a central decryption server.
Such systems can be conveniently implemented using \emph{blind decryption schemes}~\cite{Sakurai_1996}.
Blind decryption is a versatile primitive.
It can be used as a building block for many privacy critical applications,
such as privacy-preserving
payment systems~\cite{Chaum_1983}, key escrow systems, oblivious transfer protocols~\cite{Green_2007},
privacy-preserving systems for digital rights management~\cite{Perlman_2010,Lei_2012}
and private information retrieval~\cite{Schnorr_2000}.

\begin{figure}[!t]
\centering
\begin{tabular}{m{3.1cm}m{1.3cm}m{0.5cm}m{2.1cm}}
Alice & & & Encryptor\\
\includegraphics[width=0.7in]{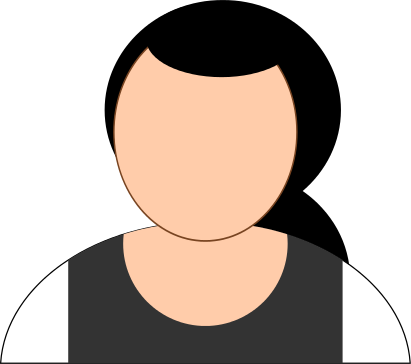}~$c_i \mapsto m_i$ & $\xleftarrow{\quad\quad\quad}$ & \[\left\{ \begin{matrix} c_1\\c_2\\ \vdots \\c_L \end{matrix} \right.\] & \includegraphics[width=0.7in]{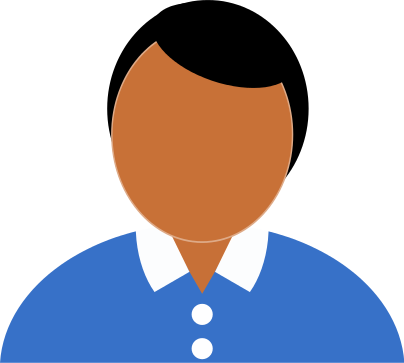} key $k$\\
~~~~~$\Big\updownarrow$ \\
\\
\includegraphics[width=0.5in]{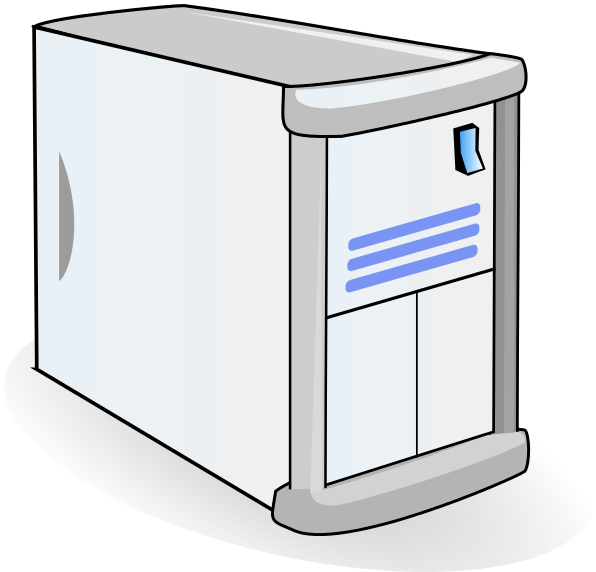}key $k$\\
Decryptor
\end{tabular}
\caption{Blind decryption. Alice has obtained $L$ ciphertexts from an encryptor and is entitled to choose exactly one of those for decryption. Alice interacts with a decryptor that shares a key $k$
with the encryptor to transform
the ciphertext message $c_i$ into a plaintext message $m_i$. Neither the encryptor nor the decryptor learn the plaintext message chosen by Alice.}
\label{fig:blind_decryption_basic}
\end{figure}

A blind decryption scheme consists of an encryption scheme together with a blind decryption protocol
intended to decrypt messages in a privacy-preserving fashion.
The meaning of ''blind decryption'' can be easily described based on the following scenario depicted in \figurename~\ref{fig:blind_decryption_basic}.
Suppose that Alice has obtained several encrypted messages from an encryptor. Alice is entitled to choose and decrypt exactly one of those messages.
Suppose that the decryption key $k$ is stored on a decryption server and Alice wishes to have the server decrypt the message for her
in such a way that neither the encryptor nor the decryptor learn the message chosen by Alice.

There are suggestions for practical blind decryption based on public key cryptography~\cite{Sakurai_1996,Sakurai_1998,Sameshima_1998,Ogata_2009,Green_2011}.
It is also possible to implement the blind decryption functionality with other protocols
such as secure multi party computation~\cite{Yao_1986}.
However, the resulting schemes would
be computationally demanding.
For many applications symmetric primitives are sufficient and
computationally more efficient.
In addition, they can provide secrecy that is not based on computational assumptions.
Oblivious transfer schemes~\cite{Rabin_1981,Even_1985} deliver the same functionality
directly between the sender and the receiver without the decryption server.
However, for noiseless channels, information-theoretically secure oblivious transfer is impossible~\cite{Damgard_1999}.
In addition, there does not seem to exist
blind decryption schemes such that the privacy of the user is based on information-theoretic security.
Our work aims to fill this shortage.
In this paper, we give a meaningful definition of perfect secrecy
for the blind decryption scenario.
In particular, we formulate
perfect secrecy of symmetric blind decryption
in a setting where at most one of the participants is maliciously observing
but adhering to the protocol.
We also propose
a
symmetric key blind decryption scheme $\sch{SymmetricBlind}$ that satisfies
our definition.
The scheme is based on modular arithmetic on a ring $\Z_{p^2}$, where $p$
is a prime.

The paper is organized as follows.
In Section~\ref{sec:Related work}, we describe work that is related to ours.
 Section~\ref{sec:Preliminaries} discusses the
fundamental definitions and the preliminaries for the rest of the paper.
In Section~\ref{sec:Perfect secrecy for symmetric blind decryption}, we formulate
three perfect secrecy properties that
the blind decryption scheme needs to satisfy.
In Section~\ref{sec:Perfectly secure encryption},
we give a description of a symmetric blind decryption scheme $\sch{SymmetricBlind}$.
In Section~\ref{sec:Security of devised scheme}, we
show that the devised scheme
satisfies our definition of perfect secrecy.
Finally, Section~\ref{sec:Future work}
considers future work and Section~\ref{sec:Conclusion}
provides the conclusion.

\section{Related work}
\label{sec:Related work}

Chaum was the first to consider blindness in the context of digital signatures
and privacy preserving payment systems~\cite{Chaum_1983}. He described the
first public key blind signature scheme~\cite{Chaum_1985} by utilizing the properties of RSA encryption~\cite{Rivest_1978}.
The scheme can be also used for encryption and can be therefore considered as the first blind decryption scheme.
In the early articles, blind decryption is referred to as ''blind decoding''.
Discrete logarithm based blind signature schemes were suggested in~\cite{Chaum_1992,Okamoto_1992,Horster_1995,Camenisch_1995}.
Sakurai and Yamane were the first to consider public key
blind decryption based on the discrete logarithm problem~\cite{Sakurai_1996}.
Their method was based on the ElGamal cryptosystem~\cite{ElGamal_1985}
and related to the blind signature of Camenisch, Piveteau and Stadler~\cite{Camenisch_1995}. The method was later applied for
the implementation of a key escrow system~\cite{Sakurai_1998}.
Mambo, Sakurai and Okamoto were the first to consider blind decryption that is secure against chosen
plaintext attacks by signing the ciphertext messages~\cite{Mambo_1996}. The resulting scheme
is not capable of public key encryption since a secret signing key is required.
Green described the first public key blind decryption scheme~\cite{Green_2011} that is secure against
adaptive chosen ciphertext attacks (IND-CCA2) using bilinear groups.
The security of these constructions has been considered computationally
either in the random oracle model~\cite{Schnorr_2000} or using computational indistinguishability
and infeasibility assumptions~\cite{Green_2011}.

\emph{Oblivious transfer} protocols are symmetric primitives 
that offer functionality similar to blind decryption.
For oblivious transfer, there are two participants: a sender and
a receiver. For the original definition of oblivious transfer, the sender transmits a message which
the receiver gets with probability $1/2$. The sender remains oblivious whether the receiver actually
got the message. This form of oblivious transfer was introduced by Rabin~\cite{Rabin_1981}. The 
concept was later extended by Even, Goldreich and Lempel~\cite{Even_1985}.
For ${2 \choose 1}$-oblivious transfer, the receiver can choose one from two messages without the sender
knowing which of the messages were chosen.
A related concept that can be considered as a further generalization
is \emph{all-or-nothing disclosure of secrets}~\cite{Brassard_1987}
for which Alice is willing to disclose at most one secret from a set
to Bob without Bob learning information about the rest of the secrets. Alice must not learn which secret Bob chose.

Adaptive queries were considered by Naor and Pinkas~\cite{Naor_1999_2}. They also considered active adversaries
and provided security definitions relating to the simulatability of the receivers.
Camenisch, Neven and Shelat extended the work of Naor and Pinkas by defining \emph{simulatable} oblivious transfer~\cite{Camenisch_2007}
and providing practical constructions for such a scheme.
There are other suggestions for oblivious transfer based on problems in bilinear groups~\cite{Green_2008},
groups of composite order~\cite{Jarecki_2009} and the Diffie-Hellman problem~\cite{Kurosawa_2009,Kurosawa_2010,Green_2011_2,Kurosawa_2011,Zhang_2013,Guleria_2015}.
These schemes
are based on computational assumptions.
It is impossible to achieve information-theoretic
security for both of the parties using noiseless channels~\cite{Damgard_1999}. However, it
is possible using noisy channels such as discrete memoryless channels~\cite{Crepeau_2005}
or a trusted initializer~\cite{Rivest_1999}.

General \emph{multiparty computation} protocols can be also applied to implement blind decryption capabilities.
Secure multiparty computation was originally introduced by Yao~\cite{Yao_1982_2} for two party case.
The general case for $n \geq 2$ is due to Goldreich, Micali and Wigderson~\cite{Goldreich_1987}.
However, secure multiparty computation protocols are computationally intensive in comparison
to pure blind decryption and oblivious transfer.

\section{Preliminaries}
\label{sec:Preliminaries}

\subsection{Notation}

For the set of integers modulo $n$,
we denote $\Z_n = \{[0],[1],\ldots,[n-1]\}$ and equate a congruence class
with its least non-negative representative. That is,
we consider $\Z_n = \{0,1,\ldots,n-1\}$. By the notation $x \bmod{n}$ we mean
the unique $i \in \{0,1,\ldots,n-1\}$ such that $i \equiv x \pmod{n}$.

We denote the uniform distribution on a set $X$ by $U(X)$. If a random variable
$\rv{Z}$ is uniformly distributed on a set $X$, we denote it by $\rv{Z} \sim U(X)$.
When an element $x$ is sampled from $U(X)$, we denote it by $x \leftarrow U(X)$.

\subsection{Symmetric encryption}

A symmetric encryption scheme $\sch{SE} = (\alg{Gen},\alg{Enc},\alg{Dec})$
with keyspace $\mathcal{K}$, plaintext space $\mathcal{M}$ and ciphertext space $\mathcal{C}$
consists of three algorithms:
\begin{enumerate}
	\item The key generation algortihm $\alg{Gen}(s)$: On input a security parameter $s$, $\alg{Gen}$ outputs a key $k \in \mathcal{K}$.
	\item The encryption algorithm $\alg{Enc}(k,m)$: On input a key $k \in \mathcal{K}$ and a message $m \in \mathcal{M}$, $\alg{Enc}$ outputs a ciphertext $c \in \mathcal{C}$.
	\item The decryption algorithm $\alg{Dec}(k,m)$: On input a key $k \in \mathcal{K}$ and a ciphertext $c \in \mathcal{C}$, $\alg{Dec}$ outputs a message $m \in \mathcal{M}$
	such that $m = \alg{Dec}(k,\alg{Enc}(k,m))$.
\end{enumerate}


\subsection{Blind decryption}
\label{sec:Blind decryption}

Blind decryption has been considered in the literature for the asymmetric case.
However, in this paper we are interested in the symmetric case which is
easily adapted from the asymmetric one~\cite{Green_2011}.
A symmetric blind decryption scheme $\sch{BlindDecryption}$
consists of a symmetric
encryption scheme $\sch{SE} = (\alg{Gen},\alg{Enc},\alg{Dec})$
	and a two-party protocol $\alg{BlindDec}$.
The protocol $\alg{BlindDec}$ is conducted between an honest user Alice and the decryption server which we shall call the Decryptor. The protocol
enables Alice, that is in possession of a ciphertext $c$, to finish the protocol
with the correct decryption of $c$.	
	As a result of running $\alg{BlindDec}$, Alice on input a ciphertext $c = \alg{Enc}(k,m) \in \mathcal{C}$
	outputs either the message $m \in \mathcal{M}$ or an error message $\perp$.
  The Decryptor, on input the key $k \in \mathcal{K}$, outputs nothing or an error message $\perp$.

 To be secure,
the exchanged messages
must not leak information to malicious users (the \emph{leak-freeness property}~\cite{Green_2007}).
The property can be formalized based on computational indistinguishability.
For every adversary, there has to be a simulator so that the following two games are well defined.
For the first game, a probabilistic polynomial time (PPT) adversary $\alg{A}$ can choose any number $L$ of ciphertexts $c_i$ for $i \in \{1,2,\ldots,L\}$.
It is then given the correct decryptions by executing $\alg{BlindDec}$ with the Decryptor. Finally,
$\alg{A}$ outputs the plaintext message, ciphertext pairs $(m_i,c_i)$ for $i \in \{1,2,\ldots,L\}$.
For the second game,
a simulator $\alg{S}$ chooses any number $L$ of ciphertexts $c_i$ for $i \in \{1,2,\ldots,L\}$.
In this game, the plaintext messages are obtained by querying a trusted party.
$\alg{BlindDecryption}$ is \emph{leak-free} if for every PPT adversary $\alg{A}$ there is a simulator $\alg{S}$
such that for every PPT distinguisher $\alg{D}$ the probability of distinguishing between
these two games is negligible~\cite{Green_2011}.

Another important property for secure blind decryption is the \emph{blindness property}.
It formalizes the idea that the Decryptor must not learn anything
about the actual plaintext message.
This can be formalized
by giving a PPT algorithm $\alg{D}$ the possibility to choose
two ciphertexts $c_1,c_2$ and giving it oracle access
to two instances of $\alg{BlindDec}$ based on these choices. If the probability of distinguishing
these two instances is negligible for every PPT algorithm $\alg{D}$, then 
$\alg{BlindDecryption}$ satisfies \emph{ciphertext blindness}.
For a formal and rigorous definition, see for example~\cite{Green_2011}.

%
%
%

\subsection{Perfect secrecy}

The notion of perfect secrecy is due to Shannon~\cite{Shannon_1949}. Let $\sch{SE} = (\alg{Gen},\alg{Enc},\alg{Dec})$ be an encryption
scheme with keyspace $\mathcal{K}$, plaintext space $\mathcal{M}$ and ciphertext space $\mathcal{C}$.
Let $K$ denote a random variable on the keyspace
induced by $\alg{Gen}$.
$\sch{SE}$
satisfies perfect secrecy if for every random variable $M$
on the plaintext space, every plaintext $m \in \mathcal{M}$ and every
ciphertext $c \in \mathcal{C}$,
\[
\Pr \left[ M = m | c = \alg{Enc}(K, M) \right] = \Pr \left[ M = m \right].
\]
Equivalently, 
$\sch{SE}$
satisfies perfect secrecy if and only if for every random variable $M$
on the plaintext space,
every plaintext messages $m_1,m_2 \in \mathcal{M}$ and every
ciphertext $c \in \mathcal{C}$,
\begin{eqnarray}
\lefteqn{\Pr \left[ c = \alg{Enc}(K, M)| M = m_1 \right]} \nonumber\\
& & = \Pr \left[ c = \alg{Enc}(K, M)| M = m_2 \right]. \nonumber
\end{eqnarray}

\section{Perfect secrecy for symmetric blind decryption}
\label{sec:Perfect secrecy for symmetric blind decryption}

Instead of computational indistinguishability, we shall now consider secrecy of
symmetric blind decryption based on the information observed by the parties.
In the following, let $\sch{SE} = (\alg{Gen},\alg{Enc},\alg{Dec})$ together with
$\alg{BlindDec}$ be a symmetric blind decryption
scheme with keyspace $\mathcal{K}$, plaintext space $\mathcal{M}$ and ciphertext space $\mathcal{C}$.

\subsection{The scenario}


For the sake of clarity, we do not consider active adversaries. We assume that the parties adhere to
the blind decryption protocol and only observe the flow of messages (and possibly deduce information from those messages).
Active adversaries could, for example, induce errors to the protocol messages.
Such adversarial scenarios are left for future work.
In addition, we do not consider the case that the Decryptor
is colluding with either Alice or the Encryptor
against the other.
Such a case is equivalent to the oblivious transfer scenario and information-theoretic security is impossible
for noiseless channels~\cite{Damgard_1999}.
However, we note that such collusion scenarios are important for certain applications and need to be investigated in the future.
We do consider the case that the adversary is impersonating one of the parties which is a paramount
requirement for many applications.

For clarity, we also restrict to the case that Alice decrypts a single message $m \in \mathcal{M}$. Similar to the one-time pad, we assume
that a new key is derived after every decryption.
However, in our case there could be several ciphertexts
$c_1,c_2,\ldots,c_L$ encrypted under the same key.
Nevertheless, once Alice has decrypted one of the messages
we consider that particular key used and a new key and a new set of ciphertexts is generated.

\begin{figure}[!t]
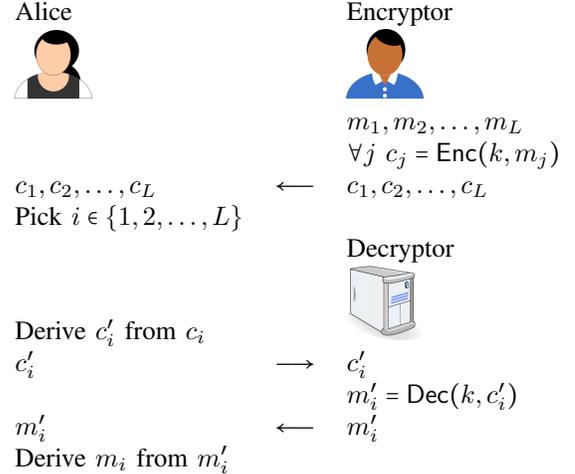

\centering
\begin{tabular}{lcl}
Alice & & Encryptor \\
\includegraphics[width=0.4in]{Alice.png} & & \includegraphics[width=0.4in]{Bob.png}\\
& & $m_1,m_2,\ldots,m_L$ \\
& & $\forall j ~ c_j = \alg{Enc}(k,m_j)$ \\
$c_1,c_2,\ldots,c_L$ & $\longleftarrow$ & $c_1,c_2,\ldots,c_L$ \\
Pick $i \in \{1,2,\ldots,L\}$ & & \\
& & Decryptor\\
Derive $c_i'$ from $c_i$ & & \includegraphics[width=0.4in]{Decryptor.png}\\
$c_i'$ & $\longrightarrow$ & $c_i'$ \\
& & $m_i' = \alg{Dec}(k,c_i')$ \\
$m_i'$ & $\longleftarrow$ & $m_i'$ \\
Derive $m_i$ from $m_i'$ & & \\
\end{tabular}
\caption{The general blind decryption scenario. Alice chooses a ciphertext $c_i$ and derives a related ciphertext $c_i'$ that she transmits to the decryptor. The decryptor responds with the corresponding plaintext message $m_i'$ from which Alice can recover $m_i$.}
\label{fig:general_scenario}
\end{figure}

The scenario is the following.
The Encryptor chooses a set of $L$ plaintext messages $m_i$ for $i \in \{1,2,\ldots,L\}$.
He encrypts those messages under a key $k$ to obtain ciphertext
messages $c_j = \alg{Enc}(k,m_j)$ for $j \in \{1,2,\ldots,L\}$ that he transmits to Alice.
Alice chooses one of those messages $c_i$.
To hide the actual ciphertext $c_i$, we assume that there is a
ciphertext transformation space $\mathcal{C}' \subseteq \mathcal{C}$ so that
Alice can derive a related ciphertext message $c_i' \in \mathcal{C}'$
that she transmits to the Decryptor. The Decryptor responds
with its decryption $m_i' \in \mathcal{M}$
which Alice transforms to the correct plaintext message $m_i$.
The general scenario has been depicted in \figurename~\ref{fig:general_scenario}.
The used variables have been collected into Table~\ref{tab:Table of symbols} for easier reference.
\begin{table}
\caption{Variables}
\begin{tabular}{|c|l|}
\hline
Symbol & Description\\
\hline
\hline
$\mathcal{K}$ & key space\\
$\mathcal{M}$ & plaintext space\\
$\mathcal{C}$ & ciphertext space\\
$\mathcal{C'}$ & ciphertext transformation space\\
$k$ & blind encryption / decryption key\\
$L$ & the number of messages encrypted under\\ & a single blind decryption key\\
$m_1,m_2,\ldots,m_L$ & plaintext messages chosen by the Encryptor\\
$c_1,c_2,\ldots,c_L$ & ciphertext messages obtained by encrypting \\ & with the blind encryption key\\
$c$ or $c_i$ & ciphertext message chosen by Alice\\
$c'$ or $c_i'$ & transformed ciphertext message chosen by Alice\\
$m'$ or $m_i'$ & decryption of $c'$ under the blind decryption key\\
$m$ or $m_i$ & the plaintext message Alice obtains at the end\\ & of the scheme\\
\hline
\end{tabular}
\vspace{4pt}
\label{tab:Table of symbols}
\end{table}

%
%
%

\subsection{Security requirements}

\begin{figure}[!t]
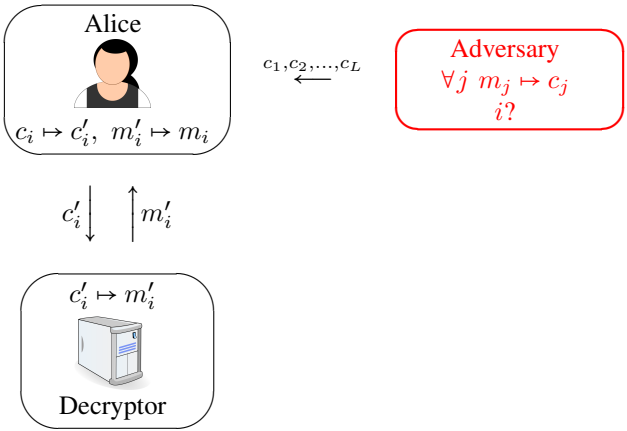

\centering
\begin{tabular}{ccc}
\ovalbox{%
\begin{minipage}{0.3\columnwidth}
\centering
Alice\\
\vspace{0.1cm}
\includegraphics[width=0.4in]{Alice.png}\\
$c_i \mapsto c_i', ~ m_i' \mapsto m_i$
\end{minipage}
}
& $\overset{c_1,c_2,\ldots,c_L}{\longleftarrow}$ & 
\color{red}
\Ovalbox{%
\begin{minipage}{0.3\columnwidth}
\centering
Adversary\\
$\forall j~m_j \mapsto c_j$\\
$i$?
\end{minipage}
}\\
 & & \\
$c_i' \Big\downarrow \quad \Big\uparrow m_i'$ & & \\
 & & \\
\ovalbox{%
\begin{minipage}{0.25\columnwidth}
\centering
$c_i' \mapsto m_i'$\\
\vspace{0.1cm}
\includegraphics[width=0.4in]{Decryptor.png}\\
\vspace{-0.1cm}
Decryptor\\
\end{minipage}
}
\end{tabular}
\caption{Malicious Encryptor. The adversary attempts to learn which message was chosen by Alice.}
\label{fig:malicious_Bob}
\end{figure}

\begin{figure}[!t]
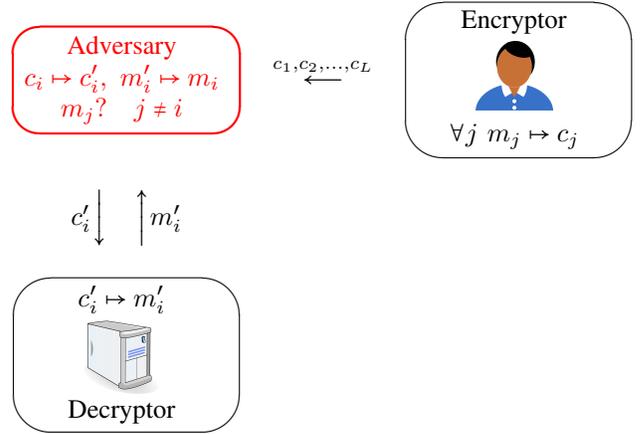

\centering
\begin{tabular}{ccc}
\color{red}
\Ovalbox{%
\begin{minipage}{0.3\columnwidth}
\centering
Adversary\\
$c_i \mapsto c_i', ~ m_i' \mapsto m_i$\\
$m_j? \quad j \neq i$
\end{minipage}
}
& $\overset{c_1,c_2,\ldots,c_L}{\longleftarrow}$ & 
\ovalbox{%
\begin{minipage}{0.3\columnwidth}
\centering
Encryptor\\
\vspace{0.1cm}
\includegraphics[width=0.4in]{Bob.png}\\
$\forall j~m_j \mapsto c_j$
\end{minipage}
}\\
 & & \\
$c_i' \Big\downarrow \quad \Big\uparrow m_i'$ & & \\
 & & \\
\ovalbox{%
\begin{minipage}{0.3\columnwidth}
\centering
$c_i' \mapsto m_i'$\\
\vspace{0.1cm}
\includegraphics[width=0.4in]{Decryptor.png}\\
\vspace{-0.1cm}
Decryptor\\
\end{minipage}
}
\end{tabular}
\caption{Malicious Alice. The adversary attempts to decrypt additional messages.}
\label{fig:adversary_alice}
\end{figure}

As described in Section~\ref{sec:Blind decryption}, the scheme has to satisfy the following property.
\begin{enumerate}
	\item Leak-freeness. Malicious observers
	must not learn information about the plaintext messages by observing the exchanges.
\end{enumerate}
The easiest way to provide leak-freeness against malicious observers
that are not participants of the scheme
is to
protect each exchange
with an encryption scheme that satisfies perfect secrecy.
However, leakage need to
be also addressed considering maliciousness of the protocol participants.
Considering each individual party, we can divide leak-freeness as follows.
\begin{enumerate}[label*=1.\arabic*)]
	\item Leak-freeness against the Encryptor. Malicious encryptor
	must not learn information about the plaintext message obtained by Alice at the end of the protocol
	by observing the blind decryption messages. The situation is depicted in
\figurename~\ref{fig:malicious_Bob}.
	\item Leak-freeness against Alice. This property ensures that, after obtaining $m_i$, Alice does not learn information about
	the remaining $L-1$ plaintexts $m_j$ for $j \neq i$. The situation is depicted in \figurename~\ref{fig:adversary_alice}.
\end{enumerate}
In contrast to computational security, we cannot define
leak-freeness as a distinguishing problem. Instead, we shall consider
the probability distributions regarding the exchanged elements.

\begin{figure}[!t]
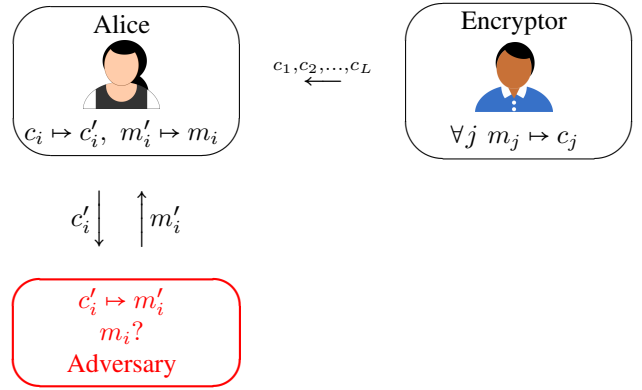

\centering
\begin{tabular}{ccc}
\ovalbox{%
\begin{minipage}{0.3\columnwidth}
\centering
Alice\\
\vspace{0.1cm}
\includegraphics[width=0.4in]{Alice.png}\\
$c_i \mapsto c_i', ~ m_i' \mapsto m_i$
\end{minipage}
}
& $\overset{c_1,c_2,\ldots,c_L}{\longleftarrow}$ & 
\ovalbox{%
\begin{minipage}{0.3\columnwidth}
\centering
Encryptor\\
\vspace{0.1cm}
\includegraphics[width=0.4in]{Bob.png}\\
$\forall j~m_j \mapsto c_j$
\end{minipage}
}\\
 & & \\
$c_i' \Big\downarrow \quad \Big\uparrow m_i'$ & & \\
 & & \\
\color{red}
\Ovalbox{%
\begin{minipage}{0.3\columnwidth}
\centering
$c_i' \mapsto m_i'$\\
$m_i?$\\
Adversary
\end{minipage}
}
\end{tabular}
\caption{Malicious Decryptor. The adversary attempts to learn the plaintext message that Alice obtains.}
\label{fig:malicious_server}
\end{figure}

We also want to prevent Decryptor from deducing information about the plaintext message $m_i$.
\begin{enumerate}
\setcounter{enumi}{1}
	\item Blindness against the Decryptor. This property ensures that a malicious decryption server does not learn the message Alice wants to decrypt. The situation
is depicted in \figurename~\ref{fig:malicious_server}.
\end{enumerate}
In the computational security setting, there can be multiple applications of the blind decryption
protocol for a fixed key.
In our case, we want a fresh key for every decryption to achieve perfect secrecy.
Therefore, we formulate leak-freeness and blindness for a single decryption.
However, as was described before, we want to be able to encrypt multiple messages with the same key.
For example, in privacy-preserving payment
systems blind decryption is used to enable Alice to choose one -- but only one -- item from a selection of items.
This results in
a scenario in which there are $L$ plaintext, ciphertext pairs $(m_j,c_j)$ for $j \in \{1,2,\ldots,L\}$
but there is only a single application of $\sch{BlindDec}$.

In the following section, we formulate these conditions based on information. Note that these
conditions also provide secrecy against malicious observers that are not participants of the
scheme since the information possessed by such observers is a proper subset of that of any of the participants.
The following
notation is used.
Let $K$
denote the random variable of blind decryption keys on the key space $\mathcal{K}$
induced by $\alg{Gen}$.
Let $M_j$ for $j \in \{1,2,\ldots,L\}$ denote
the random variables corresponding to the choice of $m_i$ for $j \in \{1,2,\ldots,L\}$
by the Encryptor and let $M$ denote the random variable corresponding to the plaintext $m$
Alice obtains at the end of the scheme. Following the standard practice~\cite{Katz_2007}, we assume that $K$ is independent
with $M$ and $M_j$ for every $j \in \{1,2,\ldots,L\}$.
Let $C'$ denote the random variable on the ciphertext transformation space $\mathcal{C'}$
for the ciphertext message $c'$ that Alice discloses to the Decryptor.
Finally, let $M'$ denote the random variable corresponding to the message $m'$
that the Decryptor responds with. These variables have been collected into Table~\ref{tab:Table of random variables}.
\begin{table}
\caption{Random variables}
\begin{tabular}{|c|l|}
\hline
Random variable & Description\\
\hline
\hline
$K$ & random variable on $\mathcal{K}$ induced by $\alg{Gen}$ \\
$M_1,M_2,\ldots,M_L$ & random variables corresponding to the choice\\ & of $m_1,m_2,\ldots,m_L$ by the encryptor\\
$C'$ & random variable on $\mathcal{C'}$ induced by Alice\\ & using $\alg{BlindDec}$\\
$M'$ & random variable on $\mathcal{M}$ induced by decryption\\ & of $C'$ by the decryptor\\
$M$ & random variable corresponding to the plaintext\\ &  message $m$ Alice obtains at the end of the scheme\\
\hline
\end{tabular}
\vspace{4pt}
\label{tab:Table of random variables}
\end{table}
%
%

\subsection{Perfect leak-freeness against the encryptor}

We shall first formulate leak-freeness against the Encryptor.
The blind decryption protocol messages $c'$ and $m'$
should not disclose any information about $m_i$ to the Encryptor. Equivalently,
the messages should not leak information about the $i$ that
was chosen by Alice even if the Encryptor knows the key $k$ and the right
plaintext messages $m_j$ for $j \in \{1,2,\ldots,L\}$.
%
%
\begin{definition}[Perfect leak-freeness against encryptor]
\label{def:leak-freeness_against_encryptor}
A symmetric blind decryption
scheme
is \emph{perfectly leak-free against the encryptor} for a single decryption
of a maximum of $L$ messages
if for every random variable $M,M_j$ for $j \in \{1,2,\ldots,L\}$ on the plaintext space
and every $m,m',m_j \in \mathcal{M}$ for $j \in \{1,2,\ldots,L\}$ and every $c' \in \mathcal{C'}$,
\begin{IEEEeqnarray}{l}
\Pr \left[ M = m \left| C' = c' \cap M' = m' \bigcap_{j = 1}^L M_j = m_j \right. \right] \nonumber\\
\quad\quad = \Pr \left[ M = m \left| \bigcap_{j = 1}^L M_j = m_j \right. \right]. \nonumber
\end{IEEEeqnarray}
\end{definition}
Our definition states that a malicious Encryptor can equally easily
guess the plaintext message Alice wanted to be decrypted with or without information
provided by the blind decryption protocol messages $c'$ and $m'$.
Note that, in the normal scenario, $M = M_i$ for some $i \in \{1,2,\ldots,L\}$.
However, we do not want to restrict the definition to such a case. For example, there could be homomorphic
blind decryption schemes
for which certain operations could be permitted on the ciphertexts.
Note also
that the Encryptor inherently possesses more information about $m$ than an outsider
since $m$ is dependent on $m_1,m_2, \ldots, m_L$.


\subsection{Perfect leak-freeness against Alice}

In order to be practical, the scheme needs to ensure that Alice is not able to decrypt messages.
Therefore, we need to ensure that 
Alice obtains neither the
decryption key nor any information about the decryptions of $c_1,c_2,\ldots,c_L$ without interacting with
the Decryptor. In addition, after a single application of $\alg{BlindDec}$, Alice must not have
any information about the remaining $L-1$ messages.
To make the requirement precise, we require that the observation of a single 
plaintext, ciphertext pair $(m_1,c_1)$
does not leak any information about the decryption of another ciphertext $c_2$.
The property is, in fact, a property of the encryption scheme.
\begin{definition}[Perfect leak-freeness against Alice]
\label{def:Perfect leak-freeness against Alice}
A symmetric encryption scheme $\sch{SE}$ satisfies \emph{perfect leak-freeness against Alice}
for a single decryption
if
for every 
random variable $M_1,M_2$ on the plaintext space, every
$m_1,m_2,m \in \mathcal{M}$ and every $c_1,c_2 \in \mathcal{C}$ such that $c_1 \neq c_2$,
\begin{IEEEeqnarray}{c}
\Pr \left[ c_1 = \alg{Enc}(K,M_1)  \cap c_2 = \alg{Enc}(K,M_2) \right. \nonumber\\
\left. \left| M_1 = m_1 \cap M_2 = m_2 \right. \right] \nonumber\\
= \Pr \left[ c_1 = \alg{Enc}(K,M_1) \cap c_2 = \alg{Enc}(K,M_2) \right. \nonumber\\
\left. \left| M_1 = m_1 \cap M_2 = m \right. \right]. \nonumber
\end{IEEEeqnarray}
\end{definition}
The condition states that
the probability of obtaining
the ciphertext pair $(c_1,c_2)$ is the same whether we encrypt $(m_1,m_2)$ or $(m_1,m)$.
%
That is, observation of the ciphertexts $c_1,c_2$
does not yield information about the decryption of $c_2$ even if we know the decryption of $c_1$.
%

\subsection{Perfect blindness against the decryptor}

We still need to consider privacy against a malicious Decryptor.
It is reasonable to assume that $c_1,c_2,\ldots,c_L$ have been delivered to Alice
using a private channel.
If the Decryptor can observe $c_j$ for $j \in \{1,2,\ldots,L\}$, it means that
he knows the corresponding plaintext messages since he is
in possession of the blind decryption key.
Therefore,
it is natural to
require that the ciphertexts are protected by a separate secure channel
between Alice and the Encryptor.
For the blindness property we want the server to learn nothing of the actual message $m$ that Alice
derives at the end of the blind decryption scheme. 
In this case, the Decryptor knows the correct key $k$
as well as the messages $c'$ and $m'$ exchanged with Alice.

%
%
\begin{definition}[Perfect ciphertext blindness against the decryptor]
\label{def:Perfect ciphertext blindness against decryptor}
A symmetric blind decryption
scheme
satisfies \emph{perfect ciphertext blindness against the decryptor}
if for every random variable $M$ on the plaintext space
and every $m,m' \in \mathcal{M}$ and every $c' \in \mathcal{C'}$
\begin{eqnarray}
\Pr \left[ M = m \left| C' = c' \cap M' = m' \right. \right] 
= \Pr \left[ M = m \right]. \nonumber
\end{eqnarray}
\end{definition}
The condition states that it is equally easy to guess the correct plaintext message with
and without the information possessed by the decryptor. Note that we have
assumed that $c_1,c_1,\ldots,c_L$ have been delivered to Alice in perfect secrecy.

\subsection{Perfect secrecy for symmetric blind decryption}

Finally, we can state our definition of perfect secrecy based on the properties defined above.
\begin{definition}[Perfect secrecy of blind decryption]
\label{def:Perfect secrecy of blind decryption}
A symmetric blind decryption
scheme consisting of a symmetric encryption scheme $\sch{SE}$ and a blind decryption protocol $\alg{BlindDec}$
satisfies perfect secrecy for symmetric blind decryption
for a single decryption of a maximum of $L$ messages
against a single malicious party
if the scheme is perfectly leak-free against the encryptor for a maximum of $L$ messages,
$\sch{SE}$ is leak-free against Alice 
and the scheme satisfies perfect ciphertext blindness against the decryptor.
\end{definition}

\section{A concrete blind decryption scheme}
\label{sec:Perfectly secure encryption}

\begin{figure}[!t]
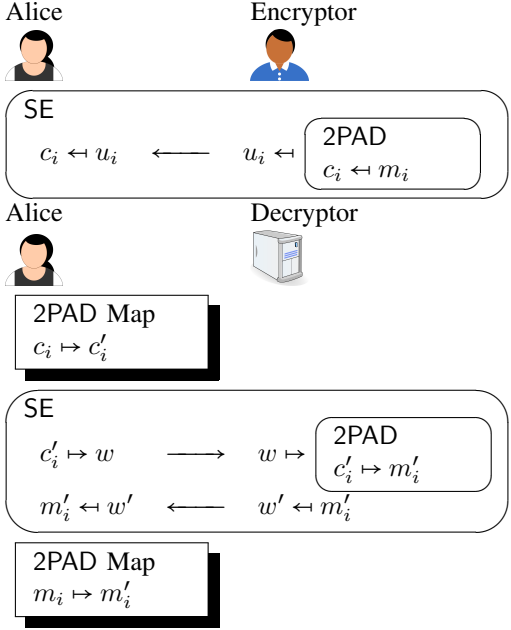

\centering
\begin{tabular}{ll}
Alice & Encryptor\\
\includegraphics[width=0.3in]{Alice.png} & \includegraphics[width=0.3in]{Bob.png}\\
\multicolumn{2}{c}{
\ovalbox{
\begin{minipage}{0.7\columnwidth}
$\sch{SE}$\\
\begin{tabular}{lcr}
$c_i \leftmapsto u_i$ & $\xleftarrow{~~~~~}$ & $u_i \leftmapsto$
\ovalbox{
\begin{minipage}{0.3\columnwidth}
$\sch{2PAD}$\\
$c_i \leftmapsto m_i$
\end{minipage}
}
\end{tabular}
\end{minipage}
}
}\\
Alice & Decryptor\\
\includegraphics[width=0.3in]{Alice.png} & \includegraphics[width=0.3in]{Decryptor.png} \\
\rule{0pt}{4ex}
\shadowbox{
\begin{minipage}{0.25\columnwidth}
$\sch{2PAD}$ Map\\
$c_i \mapsto c_i'$
\end{minipage}} & \\
\multicolumn{2}{c}{
\ovalbox{
\begin{minipage}{0.7\columnwidth}
$\sch{SE}$\\
\begin{tabular}{lll}
$c_i' \mapsto w$ & $\xrightarrow{~~~~~}$ & $w \mapsto$
\ovalbox{
\begin{minipage}{0.3\columnwidth}
$\sch{2PAD}$\\
$c_i' \mapsto m_i'$
\end{minipage}
}\\
$m_i' \leftmapsto w'$ & $\xleftarrow{~~~~~}$ & $w' \leftmapsto m_i'$
\end{tabular}
\end{minipage}
}
}\\
\rule{0pt}{8ex}
\shadowbox{
\begin{minipage}{0.25\columnwidth}
$\sch{2PAD}$ Map\\
$m_i \mapsto m_i'$
\end{minipage}}
\end{tabular}
\caption{General overview of $\sch{SymmetricBlind}$. Two tiers of encryption are applied. The outer tier ($\sch{SE}$) satisfies ordinary perfect secrecy. The inner tier ($\sch{2PAD}$) provides perfect leak-freeness against Alice and has a transformation property enabling perfect blindness against the decryptor.}
\label{fig:General overview}
\end{figure}

We shall now devise a blind decryption scheme $\sch{SymmetricBlind}$
that satisfies Def.~\ref{def:Perfect secrecy of blind decryption}.
We shall implement our scheme using two tiers of symmetric encryption.
For the outer tier we apply a scheme that satisfies
ordinary perfect secrecy.
Let that scheme be denoted by $\sch{SE}$.
The outer encryption scheme will
hide information about $c_1,c_2,\ldots,c_L$ from the Decryptor and
also provide secrecy for $c'$ and $m'$ against the Encryptor. To achieve perfect blindness
and leak-freeness against Alice, we
design an inner tier encryption scheme called $\sch{2PAD}$ that 
satisfies a useful transformation property. The property enables us to construct a blind
decryption protocol $\sch{BlindDec}$.
To sum it up, our final construction
will consist of two tiers of encryption and a protocol for Alice to query a single decryption
from the Decryptor. The general overview of the scheme is depicted in \figurename~\ref{fig:General overview}.

%

It would be possible to implement some of the required privacy properties
with multiple applications of the one time pad.
For example, if $c_i = m_i \oplus k_i$, Alice could hide
the plaintext message from the Decryptor by querying for the decryption of $c_i' = c_i \oplus k'$, where $k'$ is only known to Alice.
The correct plaintext message would be obtained from $m_i' = c_i' \oplus k_i = c_i \oplus k' \oplus k_i$
by computing $m_i' \oplus k' = c_i \oplus k_i = m_i$.
However, such a protocol would leak $i$ to the Decryptor since $i$ would be needed for decryption. In addition,
for a single decryption, the Decryptor would have to maintain a set of $L$
keys which would quickly grow to an unmanageable size as $L$ grows.
In contrast, the optimal key size for single decryption would be $2|m_i|$, where $|m_i|$ is
the bit length of $m_i$, assuming that each
plaintext message is of the same bit length.
Therefore, simply applying the one time pad is not sufficient.

%

In the following, we first describe our inner encryption scheme $\sch{2PAD}$ that will provide perfect leak-freeness against Alice,
as well as the required message transformation property.
Then, we proceed to the description of a blind decryption protocol utilizing this scheme. 
Finally, we combine the inner encryption scheme with an outer encryption scheme that satisfies
ordinary perfect secrecy and describe the complete blind decryption scheme.

\subsection{The inner encryption scheme}

We shall first construct an inner encryption scheme called $\sch{2PAD}$
with some useful properties.
Our inner scheme is based on modular arithmetic on the ring $\Z_{p^2}$, where $p \geq 5$
is a prime. Our plaintext space is $\Z_{p}$ and every $m \in \Z_{p}$ is mapped to 
$\Z_{p^2}$ which is the ciphertext space.
To satisfy Def.~\ref{def:Perfect leak-freeness against Alice}, we want to add an amount
of randomness that is at least twice the binary length of $m$ in the encryption operation.
Therefore, the keys of $\sch{2PAD}$ will consist of a pair $(x_k,y_k) \in \Z_p \times \Z_p$.

Let $z \in \Z_{p^2}$. Then,
\[
z \equiv pz' + z'' \pmod{p^2}
\]
where $z', z'' \in \Z_p$.
Therefore, we can essentially represent $z$ with two elements of $\Z_p$.
Using such a representation, we encrypt a single message $m \in \Z_p$ by
first sampling a random element $z \leftarrow U(\Z_p \setminus \{0\})$ and
setting
$b := (p m + z) \bmod{p^2}$. Then, we add the key $(x_k,y_k)$ by computing
\[
c := (p x_{k} b^2 + p y_{k} b + b) \bmod{p^2} = p x_{k} z^2 + p y_{k} z + pm + z
\]
which is the ciphertext message. Such an encryption operation
entails a useful transformation property.
For every $x_{k},y_{k} \in \Z_p$ and $b,b' \in \Z_{p^2}$ such that $b \equiv b' \pmod{p}$,
\[
p x_{k} b'^2 + p y_{k} b' + b' \equiv p x_{k} b^2 + p y_{k} b + b' \pmod{p^2}.
\]
Namely, if we know a plaintext $m_1$ and its encryption
$c_1 = p x_{k} z^2 + p y_{k} z + p m_1 + z$,
we know the decryption $m_2$ of $c_2$
for every $c_2 \equiv c_1 \pmod{p}$ since it can be computed by the following algorithm.
\begin{algorithmic}[1]
  \Procedure{$\alg{Map}$}{$c_1,m_1,c_2$}
		\State If $c_1 \not \equiv c_2 \pmod{p}$ output $\perp$
		\State $m_2 := (c_2 - c_1 + p m_1) / p$
  \State \textbf{output} $m_2$
  \EndProcedure
\end{algorithmic}
Let $z \equiv c_1 \equiv c_2 \pmod{p}$.
The algorithm works because
\begin{IEEEeqnarray}{lll}
(c_2 - c_1 + p m_1) / p & {}={} & (p x_{k} z^2 + p y_{k} z + p m_2 + z \nonumber\\
& & {-}\:p x_{k} z^2 - p y_{k} z - p m_1 - z + p m_1) / p\nonumber\\
& {}={} & (p m_2) / p \nonumber\\
& {}={} & m_2. \nonumber
\end{IEEEeqnarray}
The $\alg{Map}$ algorithm can transform the decryption $m_1$ of a ciphertext $c_1$ to the decryption $m_2$ of $c_2$ whenever $c_2 \equiv c_1 \pmod{p}$.

Decryption is straightforward knowing the key $(x_k,y_y)$. Its operation, as well as the
complete encryption scheme is described below.
\begin{definition}[$\sch{2PAD}$]
The symmetric encryption scheme
\[
\sch{2PAD} = (\alg{Gen}_{\sch{2PAD}},\alg{Enc}_{\sch{2PAD}},\alg{Dec}_{\sch{2PAD}})
\]
consists of the following three algorithms.
\begin{algorithmic}[1]
  \Procedure{$\alg{Gen}_{\sch{2PAD}}$}{$s$} \Comment{$s$ determines the size for the plaintext space}
		\State Choose a public prime $p$ such that $p \geq 5$ and $p \geq 2^s$
		\State $x_k \leftarrow U(\Z_p)$
		\State $y_k \leftarrow U(\Z_p)$
  \State \textbf{output} $(x_{k},y_{k})$
  \EndProcedure
\end{algorithmic}
\begin{algorithmic}[1]
  \Procedure{$\alg{Enc}_{\sch{2PAD}}$}{$x_{k},y_{k},m$} \Comment{Input consists of a key $(x_{k},y_{k})$ and a message $m \in \Z_{p}$}
		\State $z \leftarrow U(\Z_p \setminus \{0\})$
		\State $b := (p m + z) \bmod{p^2}$
		\State $c := (p x_{k} b^2 + p y_{k} b + b) \bmod{p^2}$
  \State \textbf{output} $c$
  \EndProcedure
\end{algorithmic}
\begin{algorithmic}[1]
  \Procedure{$\alg{Dec}_{\sch{2PAD}}$}{$x_{k},y_{k},c$} \Comment{Input consists of a key $(x_{k},y_{k})$ and a ciphertext $c \in \Z_{p^2}$}
		\State $z := c \bmod{p}$
		\State $t := (p (-x_{k}) z^2 + p (-y_{k}) z + c) \bmod{p^2}$
		\State $m := (t - z) / p$
  \State \textbf{output} $m$
  \EndProcedure
\end{algorithmic}
\end{definition}
The plaintext and ciphertext spaces of $\sch{2PAD}$ depend on the chosen prime $p$.
In particular, the plaintext space is $\Z_p$ while the ciphertext space is $\Z_{p^2}$.
Let us show the correctness of the scheme. That is,
\[
\alg{Dec}_{\sch{2PAD}}(x_{k},y_{k},\alg{Enc}_{\sch{2PAD}}(x_{k},y_{k},m)) = m
\]
for every key $(x_{k},y_{k})$ and plaintext $m$.
Let $c = \alg{Enc}_{\sch{2PAD}}(x_{k},y_{k},m)$.
Then we have
\begin{IEEEeqnarray}{lll}
c & {}={} & p x_{k} b^2 + p y_{k} b + b \nonumber\\
& {}\equiv{} & p x_{k} z^2 + p y_{k} z + p m + z \pmod{p^2} \nonumber
\end{IEEEeqnarray}
and $c \bmod p = z$, where $z \in \Z_p$. Now,
\begin{IEEEeqnarray}{lll}
\alg{Dec}_{\sch{2PAD}}(x_{k},y_{k},c)
& {}={} & (t - z) / p \nonumber\\
& {}={} & (p (-x_{k}) z^2 + p (-y_{k}) z \nonumber\\
& &  {+}\:p x_{k} z^2 + p y_{k} z + p m + z - z)/p \nonumber\\
& {}={} & (p m + z - z) / p = m.\nonumber
\end{IEEEeqnarray}

We shall later show that given a single
plaintext, ciphertext pair $(m_1,c_1)$ and a ciphertext $c_2$ such that
$c_2 \not \equiv c_1 \pmod{p}$ we still have information theoretic security for $c_2$.
That is, $\sch{2PAD}$ satisfies perfect leak-freeness against Alice whenever
$c_i \not \equiv c_j \pmod{p}$ for $i \neq j$.
However, suppose that we have two plaintext, ciphertext pairs
$(m_1,c_1),(m_2,c_2)$ such that $c_1 \not \equiv c_2 \pmod{p}$.
We can show that
the key $x_{k},y_{k}$ can be completely determined
from such two pairs.

\begin{proposition}
\label{lem:solvexy_lemma}
For every plaintext, ciphertext pair $(m_1,c_1),(m_2,c_2)$
such that $c_1 \not \equiv c_2 \pmod{p}$
there is a unique key $(x_{k}, y_{k})$ such that
\begin{IEEEeqnarray}{lll}
c_1 & {}={} & \alg{Enc}_{\sch{2PAD}}(x_{k}, y_{k}, m_1), \nonumber\\
c_2 & {}={} & \alg{Enc}_{\sch{2PAD}}(x_{k}, y_{k}, m_2). \nonumber
\end{IEEEeqnarray}
\end{proposition}
\begin{IEEEproof}
Let $z_1,z_2 \in \Z_p$ such that $z_1 \equiv c_1 \pmod{p}$ and $z_2 \equiv c_2 \pmod{p}$.
Let also $v_1 = (c_1 - p m_1 - z_1)/p$ and $v_2 = (c_2 - p m_2 - z_2)/p$.
%
Then, we have a system of
two equations
\begin{IEEEeqnarray}{lll}
v_1 & {}={} & x_{k} z_1^2 + y_{k} z_1, \nonumber\\
v_2 & {}={} & x_{k} z_2^2 + y_{k} z_2,\nonumber
\end{IEEEeqnarray}
where $v_1,v_2,z_1,z_2$ are known. Let now
\[
Z =
	\begin{pmatrix}
	z_1^2 & z_2^2 \\
	z_1 & z_2
	\end{pmatrix}.
	\]
Note that
since $z_1,z_2 \not \equiv 0 \pmod{p}$
and $z_1 \not \equiv z_2 \pmod{p}$
we have $z_1^2 z_2 - z_1 z_2^2 \not \equiv 0 \pmod{p}$ and
$Z$ is invertible modulo $p$. Therefore, the equation pair has a unique solution
\begin{IEEEeqnarray}{lll}
\begin{pmatrix}
v_1 & v_2
\end{pmatrix}
\cdot Z^{-1}
& {}={} &
\begin{pmatrix}
x_{k} z_1^2 + y_{k} z_1 & x_{k} z_2^2 + y_{k} z_2
\end{pmatrix}
\cdot Z^{-1} \nonumber\\
& {}={} &
\begin{pmatrix}
x_{k} & y_{k}
\end{pmatrix}
\begin{pmatrix}
z_1^2 & z_2^2 \\
z_1 & z_2
\end{pmatrix}
\cdot Z^{-1} \nonumber\\
& {}={} &
\begin{pmatrix}
x_{k} & y_{k}
\end{pmatrix}. \nonumber
\end{IEEEeqnarray}

\end{IEEEproof}

Due to $\alg{Map}$, we require that if Bob sends $L$ ciphertext messages
$c_1,c_2,\ldots,c_L$
to Alice we have $c_i \not \equiv c_j \pmod{p}$ for every $i \neq j$.
Therefore, the maximum number of ciphertext messages under the same key is determined by $L \leq p-1$.

\subsection{Blind decryption protocol}

Next, we give a description of a blind decryption protocol based on the transformation
algorithm $\alg{Map}$.
\begin{definition}[$\sch{BlindDec}$]
Suppose that the Encryptor and the Decryptor share a key $(x_{k},y_{k}) = \alg{Gen}_{\sch{2PAD}}(s)$
intended for a single decryption by Alice.
Furthermore, let Alice have an encrypted message $c = \alg{Enc}_{\sch{2PAD}}(x_{k},y_{k},m)$
that is not known to the Decryptor.
Finally, suppose that the prime $p$ is public knowledge.
Let the protocol $\sch{BlindDec}$ be defined by the following exchange between Alice and the Decryptor:
\begin{enumerate}
	\item Alice: Compute $c' := c \bmod{p}$ and transmit it to the Decryptor.
	\item Decryptor: Reply with
	$m' = \alg{Dec}_{\sch{2PAD}}(x_{k},y_{k},c')$.
	\item Alice: Compute the plaintext message $m = \alg{Map}(c',m',c)$.
\end{enumerate}
\end{definition}
Let us quickly check the correctness of $\sch{BlindDec}$.
Let $z \equiv c' \equiv c \pmod{p}$.
Then,
$c = p x_{k} z^2 + p y_{k} z + p m + z$, where $m$
is the plaintext message.
The Decryptor replies with
\[
m' = ( p (-x_{k}) z^2 + p (-y_{k}) z + z - z ) / p = (-x_{k}) z^2 + (-y_{k}) z.
\]
But now Alice can compute
\begin{IEEEeqnarray}{l}
\alg{Map}( c',m',c) = (c - z + p m')/p \nonumber\\
\quad\quad = (p x_{k} z^2 + p y_{k} z + p m + z - z + p m')/p \nonumber\\
\quad\quad = (p x_{k} z^2 + p y_{k} z + p m - p x_{k} z^2 - p y_{k} z)/p \nonumber\\
\quad\quad = (p m)/p \nonumber\\
\quad\quad = m \nonumber
\end{IEEEeqnarray}
which is the correct plaintext message.



\subsection{The complete blind decryption scheme}

As was mentioned earlier, the communication between Alice and the Encryptor has to be protected 
in order to prevent
the Decryptor from obtaining
the plaintext messages corresponding to $c_1,c_2,\ldots,c_L$. If the Decryptor can observe these ciphertext messages,
it can freely decrypt all them since it knows the correct key. Therefore,
we need to apply an outer encryption scheme that hides the ciphertext messages.
The same solution is
the easiest way to provide perfect leak-freeness against the Encryptor
since it enables us to simplify
the secrecy conditions.
In our case, we want to protect both of these exchanges
with an outer tier of encryption that provides perfect secrecy.
Let $\sch{SE}_n = (\alg{Gen}_n,\alg{Enc}_n,\alg{Dec}_n)$ be any symmetric encryption scheme
such that the plaintext and ciphertext space is $\Z_n$.
Let it also satisfy (ordinary) perfect secrecy.
We apply $\sch{2PAD}$ together
with $\sch{SE}_n$ to provide the required leak-freeness and blindess properties.

The outer tier is composed in the following way. Alice and the Encryptor shares
a set of keys $k_1,k_2,\ldots,k_L$. The Encryptor protects each ciphertext message
by computing $u_j = \alg{Enc}_{p^2}(k_j,c_j)$ for $j \in \{1,2,\ldots,L\}$. It sends
$u_1,u_2,\ldots,u_L$ to Alice. Similarly, Alice and the Decryptor
share a pair of keys $k_C,k_P$ that are used to protect
$c_i'$ and $m_i'$. Alice sends $w = \alg{Enc}_{p}(k_C,z)$ to the Decryptor
who responds with $w' = \alg{Enc}_p(k_P,m')$.
The resulting scheme $\sch{SymmetricBlind}$ is defined as follows.
\begin{definition}[$\sch{SymmetricBlind}$]
\label{def:the_final_scheme}
Let $\sch{SE}_n = (\alg{Gen}_n,\alg{Enc}_n,\alg{Dec}_n)$ be a symmetric encryption scheme
such that the plaintext and ciphertext space is $\Z_n$ and
let $\sch{SE}_n$ satisfy perfect secrecy.
Let Alice and the Encryptor share a set of keys $k_1,k_2,\ldots,k_L$. Let
Alice and the Decryptor share a pair of keys $k_C,k_P$ intended for
a single blind decryption by Alice. Let also the Encryptor and the Decryptor
share a blind decryption key $(x_k,y_k) = \alg{Gen}_{\sch{2PAD}}(s)$, where $2^s \geq L +1$,
that is intended for single blind decryption
by Alice. $\sch{SymmetricBlind}$ is determined by the following protocol.
\begin{center}
\begin{tabular}{|lll|}
\hline
\underline{Alice} & & \underline{Encryptor} \\
& & Choose $m_1,m_2,\ldots,m_L$ \\
& & $\forall j:$ \\ & & $c_j = \alg{Enc}_{\sch{2PAD}}(x_{k},y_{k},m_j)$ \\
& & such that \\
& & $c_{j} \not \equiv c_{j'}\pmod{p}~\forall j \neq j'$\\
& & $\forall j: ~ u_j = \alg{Enc}_{p^2}(k_j,c_j)$ \\
$u_1,u_2,\ldots,u_L$ & $\longleftarrow$ & $u_1,u_2,\ldots,u_L$ \\
$\forall j ~ c_j = \alg{Dec}_{p^2}(k_j,u_j)$ & & \\
Pick $i$ & & \\
$c' = c_i \bmod{p}$ & & \\
$w = \alg{Enc}_{p}(k_C,c')$ & & \underline{Decryptor}\\
$w$ & $\longrightarrow$ & $w$ \\
& & $c' = \alg{Dec}_p(k_C,w)$ \\
& & $m' = \alg{Dec}_{\sch{2PAD}}(x_{k},y_{k},c')$ \\
& & $w' = \alg{Enc}_p(k_P,m')$ \\
$w'$ & $\longleftarrow$ & $w'$ \\
$m' = \alg{Dec}_p(k_P,w')$ & & \\
$m_i = \alg{Map}(c',m',c_i)$ & & \\
\hline
\end{tabular}
\end{center}
\end{definition}

%

\section{Security of SymmetricBlind}
\label{sec:Security of devised scheme}

We shall now consider the security of $\sch{SymmetricBlind}$. We proceed to show that the devised
scheme satisfies the three conditions formulated in Section~\ref{sec:Perfect secrecy for symmetric blind decryption}:
perfect leak-freeness against the encryptor and Alice and perfect blindness against the decryptor.

\subsection{Perfect leak-freeness against the encryptor}

\begin{proposition}
$\sch{SymmetricBlind}$ satisfies perfect leak-freeness against the encryptor for a single decryption
of a maximum of $L \leq p-1$ messages, where $p$ is determined by $\alg{Gen}_{\sch{2PAD}}(s)$.
\end{proposition}
\begin{IEEEproof}
The claim follows directly from the observation that the Encryptor sees only $w$ and $w'$. By the description
of $\sch{SymmetricBlind}$, $c'$
and $m'$ are protected by encryption satisfying perfect secrecy and thus do not
leak information to the Encryptor.
\end{IEEEproof}

It is easy to see that the outer tier of encryption is necessary.
Suppose that the outer encryption scheme was not applied. Then $c'$
would leak $c_i \bmod{p}$ which would betray $i$ to the Encryptor.

\subsection{Perfect blindness against decryptor}

We shall now prove that the Decryptor does not get information about the plaintext message.
\begin{proposition}
$\sch{SymmetricBlind}$ satisfies perfect blindness againt the decryptor for a single blind decryption.
\end{proposition}
\begin{IEEEproof}
Since $c_1,c_2,\ldots,c_L$ are protected with perfect secrecy, we only need to show that
\begin{eqnarray}
\Pr \left[ M = m \left| C' = c' \cap M' = m' \right. \right] 
= \Pr \left[ M = m \right], \nonumber
\end{eqnarray}
where $C'$ and $M'$ are the random variables associated to the messages $c'$ and $m'$, respectively.
Let $X,Y$ denote the random variables corresponding to the key elements $(x_k,y_k) \leftarrow \alg{Gen}(s)$,
respectively.
The reply $m'$ from the Decryptor is completely determined by the key $(x_{k},y_{k})$
and the element $c' = c_i \bmod{p}$ since $m' = (-x_{k}) c'^2 + (-y_{k}) c'$. Therefore,
\begin{IEEEeqnarray}{l}
\Pr \left[ M = m \left| C' = c' \cap M' = m' \right. \right] \nonumber\\
\quad \quad = \Pr \left[ M = m \left| \rv{X} = x_{k} \cap \rv{Y} = y_{k} \cap C' = c' \right. \right]. \nonumber
\end{IEEEeqnarray}

Let us consider $C'$. By the description of the scheme, we have
$C' = C_i \bmod{p}$, where $i$ is the chosen index of Alice. But for every
$i$ we have, by the description of $\alg{Enc}_{\sch{2PAD}}$, that $C_i \bmod{p} \sim U(\Z_p \setminus \{0\})$.
Therefore, $C'$ is independent with $\rv{X}$ and $\rv{Y}$ and
\begin{IEEEeqnarray}{l}
\Pr \left[ M = m \left| \rv{X} = x_{k} \cap \rv{Y} = y_{k} \cap C' = z \right. \right] \nonumber\\
\quad \quad = \Pr \left[ M = m \left| \rv{X} = x_{k} \cap \rv{Y} = y_{k} \cap C' = z' \right. \right] \nonumber
\end{IEEEeqnarray}
for every $z,z' \in \Z_p \setminus \{0\}$
and
\begin{IEEEeqnarray}{l}
\Pr \left[ M = m \left| \rv{X} = x_{k} \cap \rv{Y} = y_{k} \right. \right] \nonumber\\
\quad \quad = \sum_{z \in \Z_p \setminus \{0\}}
\Pr \left[ M = m \left| \rv{X} = x_{k} \cap \rv{Y} = y_{k} \cap C' = z \right. \right] \nonumber\\
\quad \quad \quad \quad \quad \quad \cdot \Pr \left[  C' = z \left| \rv{X} = x_{k} \cap \rv{Y} = y_{k} \right. \right] \nonumber\\
\quad \quad = \frac{1}{p-1} \cdot \sum_{z \in \Z_p \setminus \{0\}}
\Pr \left[ M = m \left| \rv{X} = x_{k} \cap \rv{Y} = y_{k} \cap C' = z \right. \right] \nonumber\\
\quad \quad = \Pr \left[ M = m \left| \rv{X} = x_{k} \cap \rv{Y} = y_{k} \cap C' = z \right. \right] \nonumber
\end{IEEEeqnarray}
for any $z \in \Z_p$.

By our assumption, $M$ is independent with $\rv{X}$ and $\rv{Y}$ and therefore we have
\[
\Pr \left[ M = m \left| \rv{X} = x_{k} \cap \rv{Y} = y_{k} \right. \right] = \Pr \left[ M = m \right]
\]
which shows our claim.
\end{IEEEproof}

The proof shows that the Decryptor (with the knowledge of the key $(x_k,y_k)$ and $c'$ and $m'$)
does not gain any information
about the plaintext message $m$
assuming that
$c_j$ for $j \in \{1,2,\ldots,L\}$ have been delivered to Alice in perfect secrecy.
Considering the secrecy against the Decryptor, it would suffice send $c'$
without the additional level of encryption.
However, the additional level is necessary
to achieve leak-freeness against the Encryptor.

\subsection{Perfect leak-freeness against Alice}

We shall now
consider a malicious Alice and
show that the observation of a single plaintext, ciphertext pair
$(m_1,c_1)$ does not yield information about the decryption of $c_2$ for $c_2 \not \equiv c_1 \pmod{p}$.
\begin{proposition}
\label{prop:ConjE_secure}
$\sch{SymmetricBlind}$
satisfies
perfect leak-freeness against Alice for a single decryption
of a maximum of $L \leq p-1$ ciphertexts.
\end{proposition}
\begin{IEEEproof}
By the description of $\sch{SymmetricBlind}$, 
the ciphertext messages $c_1,c_2,\ldots,c_L$
are of different congruence class modulo $p$.
Let $M_1, M_2$ be random variables over the plaintext space $\Z_p$.
Let
$\rv{X},\rv{Y}$ denote the random variables corresponding to the key elements
$(x_{k},y_{k}) = \alg{Gen}_{\sch{2PAD}}(s)$.
We have to show that
\begin{IEEEeqnarray}{l}
\Pr \left[ c_1 = \alg{Enc}_{\sch{2PAD}}(\rv{X},\rv{Y},M_1) \cap c_2 = \alg{Enc}_{\sch{2PAD}}(\rv{X},\rv{Y},M_2) \right. \nonumber\\
\quad \quad \left. | M_1 = m_1 \cap M_2 = m_2 \cap c_1 \not \equiv c_2 \pmod{p} \right] \nonumber \\
=
\Pr
\left[ c_1 = \alg{Enc}_{\sch{2PAD}}(\rv{X},\rv{Y},M_1) \cap c_2 = \alg{Enc}_{\sch{2PAD}}(\rv{X},\rv{Y},M_2) \right. \nonumber\\
\quad \quad \left. | M_1 = m_1 \cap M_2 = m \cap c_1 \not \equiv c_2 \pmod{p} \right] \nonumber
\end{IEEEeqnarray}
for every $m_1,m_2,m \in \{0,1,2,\ldots,p-1\}$ and $c_1,c_2 \in \Z_{p^2}$ such that $c_1 \not \equiv c_2 \pmod{p}$.

Given a valid assignment for $m_1,c_1$ and $c_2$, it suffices to show that
\begin{IEEEeqnarray}{l}
\Pr
\left[ c_1 = \alg{Enc}_{\sch{2PAD}}(\rv{X},\rv{Y},M_1) \cap c_2 = \alg{Enc}_{\sch{2PAD}}(\rv{X},\rv{Y},M_2) \right. \nonumber\\
\quad \quad \left. | M_1 = m \cap M_2 = m_2 \cap c_1 \not \equiv c_2 \pmod{p} \right] = 1 / p^2 \nonumber
\end{IEEEeqnarray}
for every $m \in \Z_p$.
By Proposition~\ref{lem:solvexy_lemma}, for every plaintext, ciphertext pair $(m_1,c_1),(m,c_2)$
such that $c_1 \not \equiv c_2 \pmod{p}$
there is a unique key $(x_{k},y_{k})$.
%
Therefore,
\begin{IEEEeqnarray}{l}
\Pr
\left[ c_1 = \alg{Enc}_{\sch{2PAD}}(\rv{X},\rv{Y},M_1) \cap c_2 = \alg{Enc}_{\sch{2PAD}}(\rv{X},\rv{Y},M_2)\right. \nonumber\\
\quad\quad \quad\quad \left. | M_1 = m_1 \cap M_2 = m \cap c_1 \not \equiv c_2 \pmod{p} \right] \nonumber \\
\quad\quad =
\Pr
\left[ \rv{X} = x_{k} \cap \rv{Y} = y_{k} \right]. \nonumber
\end{IEEEeqnarray}
By the definition of $\alg{Gen}_{\sch{2PAD}}$, $\rv{X}$ and $\rv{Y}$ are independent
and we have 
\begin{IEEEeqnarray}{lll}
\Pr
\left[ \rv{X} = x_{k} \cap \rv{Y} = y_{k} \right]
& {}={} &
\Pr \left[ \rv{X} = x_{k} \right]
\cdot
\Pr \left[ \rv{Y} = y_{k} \right] \nonumber\\
& {}={} & 1 / p^2.\nonumber
\end{IEEEeqnarray}

\end{IEEEproof}

%
%
%

We have now established the perfect secrecy of 
$\sch{SymmetricBlind}$ according to Def.~\ref{def:Perfect secrecy of blind decryption}.

\subsection{The parameters}

An optimal encryption scheme, with plaintext space $\mathcal{M}$,
that satisfies perfect leak-freeness against Alice for a single decryption
needs
$2 \log_2 |\mathcal{M}|$ bits of randomness for a key.
$\sch{2PAD}$ achieves exactly this bound
since the plaintext space is $\Z_p$ and
a single key $(x_k,y_k)$ contains $2 \log_2 p$ bits of randomness.
Assuming that messages and keys are represented by binary strings, we
need $2 \lceil \log_2 p \rceil$ bits of key
to encrypt messages of length $\lfloor \log_2 p \rfloor$.
For a single decryption with $\sch{SymmetricBlind}$, the Decryptor needs to store the key elements $x_{k},y_{k} \in \Z_p$, as well as
the keys $k_C,k_P$.
The keys $k_C,k_P$ are used to encrypt messages of $\Z_p$. Therefore, 
$\lceil \log_2 p \rceil$ bits for each of these keys suffices for perfect secrecy.
In total, the Decryptor needs to store
key material of $4 \lceil \log_2 p \rceil$ bits
for a single decryption of a message of bit length $\lfloor \log_2 p \rfloor$.

Since the ciphertext space is $\Z_{p^2}$,
the ciphertext length in bits is approximately twice the plaintext length.
Depending on the length of the plaintext messages and the needed maximum number of encryptions $L \leq p-1$, we should therefore
choose the smallest possible $p$, since its bit size has no effect on the security of the scheme. Table~\ref{tab:Various parameter examples}
lists some possible choices for $p$ and the resulting key, plaintext and ciphertext lengths in bits. Note that
for long plaintext messages the maximum number of messages $L$ is practically unlimited.
\begin{table}%
\caption{Parameter examples for SymmetricBlind}
\begin{tabular}{ccccc}
$p$ & Decryptor key length & plaintext length & ciphertext length\\
& [bits] & [bits] & [bits]\\
\hline
5 & 12 & 3 & 5\\
7 & 12 & 3 & 6\\
11 & 16 & 4 & 7\\
23 & 20 & 5 & 10\\
101 & 28 & 7 & 14\\
1009 & 40 & 10 & 20\\
5003 & 52 & 13 & 25\\
20011 & 60 & 15 & 29\\
$2^{31} - 1$ & 124 & 31 & 62\\
$2^{61} - 1$ & 244 & 61 & 122\\
$2^{127} - 1$ & 508 & 127 & 254\\
\end{tabular}
\vspace{4pt}
\label{tab:Various parameter examples}
\end{table}

%

%

\section{Future work}
\label{sec:Future work}

There are two main drawbacks of the construction presented in this paper.
First, we have not considered active adversaries. Similar to the one time pad, we have only considered
such adversaries that observe the flow of messages. For practical scenarios, we need to consider
adversaries that actively induce errors into the protocol flow. However, such considerations
are most naturally conducted in the computational infeasibility model which has been used, for instance, in~\cite{Green_2011}.
In the active adversaries setting, it would also be
natural to consider the security of the devised scheme in the framework of computational indistinguishability such that
the truly random keys are exchanged with pseudorandom bit strings. In particular, the computationally hard version of
our scheme yields efficient practical implementations.

The second drawback is that we have only considered the case of a single malicious party. While it does not make
sense to consider a scenario where Alice is colluding with the Encryptor against the Decryptor,
the scenario where the Encryptor and the Decryptor are colluding is an important one. For many scenarios Alice cannot be certain
whether the Encryptor and the Decryptor are in fact separate entities.
However, if they are a single entity, the scenario is identical to oblivious transfer.
We cannot achieve information-theoretic security in such a case~\cite{Damgard_1999}.
For example,
it is easy to see that our construction fails for colluding Encryptor and Decryptor.
If that is the case, we effectively remove the outer layer of encryption which means
that $c' = c_i \bmod{p}$ leaks $i$ to the adversary.
To provide security against
colluding Encryptor and Decryptor, we would need to detect such collusion or to turn to computational
assumptions. We leave the question as an open problem for future research. 

Another interesting question for future work is to consider the case where we do not apply the outer layer
of encryption from the Encryptor to Alice. Thus far, we have defined perfect blindness so that the Decryptor has absolutely no information about
the plaintext message.
However, we could relax the requirement so that -- similar to leak-freeness against the encryptor --
the information is conditioned on the plaintexts $m_1,m_2,\ldots,m_L$. In other words, we could relax the requirement so
that the Decryptor may observe the selection (and the corresponding plaintext messages) given to Alice.
Such a relaxation
is natural in the oblivious transfer case where the Encryptor and the Decryptor are the same entity.
We could then define blindness as a property
requiring only that the selection $i$ is hidden. It is again easy to see that our scheme without the outer layer
of encryption fails such a property. If $c_1,c_2,\ldots,c_L$ are not protected, then $c' = c_i \bmod{p}$ leaks the selection $i$.
We leave this consideration also for future work.

\section{Conclusion}
\label{sec:Conclusion}

In this paper, we give a definition of perfect secrecy for symmetric blind decryption in the setting where
one of the parties may be malicious but adhering to the protocol of the scheme.
We neither consider active adversaries nor the setting where two of the participants
are colluding against the third.
We construct a symmetric blind decryption scheme $\sch{SymmetricBlind}$ and show that it satisfies our definition
of perfect secrecy. The scheme is based on two layers of encryption, where the inner layer utilizes a novel
encryption scheme $\sch{2PAD}$ given in this paper. $\sch{2PAD}$ is based on modular arithmetic with $\Z_{p^2}$
as the ciphertext space, $\Z_{p}$ as the plaintext space and $\Z_p \times \Z_p$ as the key space, where $p \geq 5$ is a prime.
The security of $\sch{SymmetricBlind}$ is shown information theoretically and does not depend on the size of $p$.
For a fixed blind decryption key, $\sch{SymmetricBlind}$ supports a single blind decryption from a selection
of $L \leq p-1$ messages.
For a single decryption of a message of bit length $\lfloor \log_2 p \rfloor$, the decryption server
needs to store key material of $4 \lceil \log_2 p \rceil$ bits.

%


\begin{thebibliography}{10}
\providecommand{\url}[1]{#1}
\csname url@samestyle\endcsname
\providecommand{\newblock}{\relax}
\providecommand{\bibinfo}[2]{#2}
\providecommand{\BIBentrySTDinterwordspacing}{\spaceskip=0pt\relax}
\providecommand{\BIBentryALTinterwordstretchfactor}{4}
\providecommand{\BIBentryALTinterwordspacing}{\spaceskip=\fontdimen2\font plus
\BIBentryALTinterwordstretchfactor\fontdimen3\font minus
  \fontdimen4\font\relax}
\providecommand{\BIBforeignlanguage}[2]{{%
\expandafter\ifx\csname l@#1\endcsname\relax
\typeout{** WARNING: IEEEtran.bst: No hyphenation pattern has been}%
\typeout{** loaded for the language `#1'. Using the pattern for}%
\typeout{** the default language instead.}%
\else
\language=\csname l@#1\endcsname
\fi
#2}}
\providecommand{\BIBdecl}{\relax}
\BIBdecl

\bibitem{Thuraisingham_2015}
\BIBentryALTinterwordspacing
B.~Thuraisingham, ``Big data security and privacy,'' in \emph{Proceedings of
  the 5th ACM Conference on Data and Application Security and Privacy}, ser.
  CODASPY '15.\hskip 1em plus 0.5em minus 0.4em\relax New York, NY, USA: ACM,
  2015, pp. 279--280. [Online]. Available:
  \url{http://doi.acm.org/10.1145/2699026.2699136}
\BIBentrySTDinterwordspacing

\bibitem{HIPAA}
{Office for Civil Rights, United State Department of Health and Human
  Services}, ``Medical privacy. national standards of protect the privacy of
  personal-health-information,''
  http://www.hhs.gov/ocr/privacy/hipaa/administrative/privacyrule/index.html
  (retrieved 29 April 2013).

\bibitem{DataProtDirective}
{European Parliament}, ``Directive {95/46/EC} of the {E}uropean {P}arliament
  and of the {C}ouncil of 24 october 1995 on the protection of individuals with
  regard to the processing of personal data and on the free movement of such
  data,'' http://eur-lex.europa.eu/ (retrieved 21.9.2012), 1995.

\bibitem{Coull_2009}
\BIBentryALTinterwordspacing
S.~Coull, M.~Green, and S.~Hohenberger,
  ``\BIBforeignlanguage{English}{Controlling access to an oblivious database
  using stateful anonymous credentials},'' in
  \emph{\BIBforeignlanguage{English}{Public Key Cryptography -- PKC 2009}},
  ser. Lecture Notes in Computer Science, S.~Jarecki and G.~Tsudik, Eds.\hskip
  1em plus 0.5em minus 0.4em\relax Springer Berlin Heidelberg, 2009, vol. 5443,
  pp. 501--520. [Online]. Available:
  \url{http://dx.doi.org/10.1007/978-3-642-00468-1\_28}
\BIBentrySTDinterwordspacing

\bibitem{Green_2011}
\BIBentryALTinterwordspacing
M.~Green, ``Secure blind decryption,'' in \emph{Public Key Cryptography -- PKC
  2011}, ser. Lecture Notes in Computer Science, D.~Catalano, N.~Fazio,
  R.~Gennaro, and A.~Nicolosi, Eds.\hskip 1em plus 0.5em minus 0.4em\relax
  Springer Berlin / Heidelberg, 2011, vol. 6571, pp. 265--282,
  10.1007/978-3-642-19379-8\_16. [Online]. Available:
  \url{http://dx.doi.org/10.1007/978-3-642-19379-8\_16}
\BIBentrySTDinterwordspacing

\bibitem{Sakurai_1996}
\BIBentryALTinterwordspacing
K.~Sakurai and Y.~Yamane, ``Blind decoding, blind undeniable signatures, and
  their applications to privacy protection,'' in \emph{Information Hiding},
  ser. Lecture Notes in Computer Science, R.~Anderson, Ed.\hskip 1em plus 0.5em
  minus 0.4em\relax Springer Berlin Heidelberg, 1996, vol. 1174, pp. 257--264.
  [Online]. Available: \url{http://dx.doi.org/10.1007/3-540-61996-8\_45}
\BIBentrySTDinterwordspacing

\bibitem{Chaum_1983}
\BIBentryALTinterwordspacing
D.~Chaum, ``\BIBforeignlanguage{English}{Blind signatures for untraceable
  payments},'' in \emph{\BIBforeignlanguage{English}{Advances in Cryptology}},
  D.~Chaum, R.~Rivest, and A.~Sherman, Eds.\hskip 1em plus 0.5em minus
  0.4em\relax Springer US, 1983, pp. 199--203. [Online]. Available:
  \url{http://dx.doi.org/10.1007/978-1-4757-0602-4\_18}
\BIBentrySTDinterwordspacing

\bibitem{Green_2007}
\BIBentryALTinterwordspacing
M.~Green and S.~Hohenberger, ``Blind identity-based encryption and simulatable
  oblivious transfer,'' in \emph{Advances in Cryptology -- ASIACRYPT 2007},
  ser. Lecture Notes in Computer Science, K.~Kurosawa, Ed.\hskip 1em plus 0.5em
  minus 0.4em\relax Springer Berlin Heidelberg, 2007, vol. 4833, pp. 265--282.
  [Online]. Available: \url{http://dx.doi.org/10.1007/978-3-540-76900-2\_16}
\BIBentrySTDinterwordspacing

\bibitem{Perlman_2010}
\BIBentryALTinterwordspacing
R.~Perlman, C.~Kaufman, and R.~Perlner, ``Privacy-preserving {DRM},'' in
  \emph{Proceedings of the 9th Symposium on Identity and Trust on the
  Internet}, ser. IDTRUST '10.\hskip 1em plus 0.5em minus 0.4em\relax New York,
  NY, USA: ACM, 2010, pp. 69--83. [Online]. Available:
  \url{http://doi.acm.org/10.1145/1750389.1750399}
\BIBentrySTDinterwordspacing

\bibitem{Lei_2012}
L.~L. Win, T.~Thomas, and S.~Emmanuel, ``Privacy enabled digital rights
  management without trusted third party assumption,'' \emph{Multimedia, IEEE
  Transactions on}, vol.~14, no.~3, pp. 546--554, June 2012.

\bibitem{Schnorr_2000}
C.~P. Schnorr and M.~Jakobsson, ``Security of signed {E}l{G}amal encryption,''
  in \emph{Advances in cryptology---ASIACRYPT 2000}, ser. Lecture Notes in
  Comput. Sci.\hskip 1em plus 0.5em minus 0.4em\relax Berlin: Springer, 2000,
  vol. 1976, pp. 73--89.

\bibitem{Sakurai_1998}
\BIBentryALTinterwordspacing
K.~Sakurai, Y.~Yamane, S.~Miyazaki, and T.~Inoue, ``A key escrow system with
  protecting user's privacy by blind decoding,'' in \emph{Information
  Security}, ser. Lecture Notes in Computer Science, E.~Okamoto, G.~Davida, and
  M.~Mambo, Eds.\hskip 1em plus 0.5em minus 0.4em\relax Springer Berlin
  Heidelberg, 1998, vol. 1396, pp. 147--157. [Online]. Available:
  \url{http://dx.doi.org/10.1007/BFb0030417}
\BIBentrySTDinterwordspacing

\bibitem{Sameshima_1998}
\BIBentryALTinterwordspacing
Y.~Sameshima, ``A key escrow system of the {RSA} cryptosystem,'' in
  \emph{Information Security}, ser. Lecture Notes in Computer Science,
  E.~Okamoto, G.~Davida, and M.~Mambo, Eds.\hskip 1em plus 0.5em minus
  0.4em\relax Springer Berlin Heidelberg, 1998, vol. 1396, pp. 135--146.
  [Online]. Available: \url{http://dx.doi.org/10.1007/BFb0030416}
\BIBentrySTDinterwordspacing

\bibitem{Ogata_2009}
W.~Ogata \emph{et~al.}, ``New identity-based blind signature and blind
  decryption scheme in the standard model,'' \emph{{IEICE} transactions on
  fundamentals of electronics, communications and computer sciences}, vol.~92,
  no.~8, pp. 1822--1835, 2009.

\bibitem{Yao_1986}
A.~C.-C. Yao, ``How to generate and exchange secrets,'' in \emph{Foundations of
  Computer Science, 1986., 27th Annual Symposium on}, Oct 1986, pp. 162--167.

\bibitem{Rabin_1981}
M.~O. Rabin, ``How to exchange secrets with oblivious transfer,'' Technical
  Report TR-81, Aiken Computation Lab, Harvard University, 1981.

\bibitem{Even_1985}
\BIBentryALTinterwordspacing
S.~Even, O.~Goldreich, and A.~Lempel, ``A randomized protocol for signing
  contracts,'' \emph{Commun. ACM}, vol.~28, no.~6, pp. 637--647, Jun. 1985.
  [Online]. Available: \url{http://doi.acm.org/10.1145/3812.3818}
\BIBentrySTDinterwordspacing

\bibitem{Damgard_1999}
\BIBentryALTinterwordspacing
I.~Damg{\aa}rd, J.~Kilian, and L.~Salvail, ``On the (im)possibility of basing
  oblivious transfer and bit commitment on weakened security assumptions,'' in
  \emph{Proceedings of the 17th International Conference on Theory and
  Application of Cryptographic Techniques}, ser. EUROCRYPT'99.\hskip 1em plus
  0.5em minus 0.4em\relax Berlin, Heidelberg: Springer-Verlag, 1999, pp.
  56--73. [Online]. Available:
  \url{http://dl.acm.org/citation.cfm?id=1756123.1756131}
\BIBentrySTDinterwordspacing

\bibitem{Chaum_1985}
\BIBentryALTinterwordspacing
D.~Chaum, ``Security without identification: Transaction systems to make big
  brother obsolete,'' \emph{Commun. ACM}, vol.~28, no.~10, pp. 1030--1044, Oct.
  1985. [Online]. Available: \url{http://doi.acm.org/10.1145/4372.4373}
\BIBentrySTDinterwordspacing

\bibitem{Rivest_1978}
R.~L. Rivest, A.~Shamir, and L.~Adleman, ``A method for obtaining digital
  signatures and public-key cryptosystems,'' \emph{Comm. ACM}, vol.~21, no.~2,
  pp. 120--126, 1978.

\bibitem{Chaum_1992}
\BIBentryALTinterwordspacing
D.~Chaum and T.~Pedersen, ``\BIBforeignlanguage{English}{Wallet databases with
  observers},'' in \emph{\BIBforeignlanguage{English}{Advances in Cryptology --
  CRYPTO'92}}, ser. Lecture Notes in Computer Science, E.~Brickell, Ed.\hskip
  1em plus 0.5em minus 0.4em\relax Springer Berlin Heidelberg, 1993, vol. 740,
  pp. 89--105. [Online]. Available:
  \url{http://dx.doi.org/10.1007/3-540-48071-4\_7}
\BIBentrySTDinterwordspacing

\bibitem{Okamoto_1992}
\BIBentryALTinterwordspacing
T.~Okamoto, ``\BIBforeignlanguage{English}{Provably secure and practical
  identification schemes and corresponding signature schemes},'' in
  \emph{\BIBforeignlanguage{English}{Advances in Cryptology -- CRYPTO'92}},
  ser. Lecture Notes in Computer Science, E.~Brickell, Ed.\hskip 1em plus 0.5em
  minus 0.4em\relax Springer Berlin Heidelberg, 1993, vol. 740, pp. 31--53.
  [Online]. Available: \url{http://dx.doi.org/10.1007/3-540-48071-4\_3}
\BIBentrySTDinterwordspacing

\bibitem{Horster_1995}
\BIBentryALTinterwordspacing
P.~Horster, M.~Michels, and H.~Petersen,
  ``\BIBforeignlanguage{English}{Meta-message recovery and meta-blind signature
  schemes based on the discrete logarithm problem and their applications},'' in
  \emph{\BIBforeignlanguage{English}{Advances in Cryptology -- ASIACRYPT'94}},
  ser. Lecture Notes in Computer Science, J.~Pieprzyk and R.~Safavi-Naini,
  Eds.\hskip 1em plus 0.5em minus 0.4em\relax Springer Berlin Heidelberg, 1995,
  vol. 917, pp. 224--237. [Online]. Available:
  \url{http://dx.doi.org/10.1007/BFb0000437}
\BIBentrySTDinterwordspacing

\bibitem{Camenisch_1995}
\BIBentryALTinterwordspacing
J.~Camenisch, J.-M. Piveteau, and M.~Stadler,
  ``\BIBforeignlanguage{English}{Blind signatures based on the discrete
  logarithm problem},'' in \emph{\BIBforeignlanguage{English}{Advances in
  Cryptology -- EUROCRYPT'94}}, ser. Lecture Notes in Computer Science,
  A.~De~Santis, Ed.\hskip 1em plus 0.5em minus 0.4em\relax Springer Berlin
  Heidelberg, 1995, vol. 950, pp. 428--432. [Online]. Available:
  \url{http://dx.doi.org/10.1007/BFb0053458}
\BIBentrySTDinterwordspacing

\bibitem{ElGamal_1985}
T.~ElGamal, ``A public key cryptosystem and a signature scheme based on
  discrete logarithms,'' \emph{IEEE Trans. Inform. Theory}, vol.~31, no.~4, pp.
  469--472, 1985.

\bibitem{Mambo_1996}
\BIBentryALTinterwordspacing
M.~Mambo, K.~Sakurai, and E.~Okamoto, ``\BIBforeignlanguage{English}{How to
  utilize the transformability of digital signatures for solving the oracle
  problem},'' in \emph{\BIBforeignlanguage{English}{Advances in Cryptology --
  ASIACRYPT '96}}, ser. Lecture Notes in Computer Science, K.~Kim and
  T.~Matsumoto, Eds.\hskip 1em plus 0.5em minus 0.4em\relax Springer Berlin
  Heidelberg, 1996, vol. 1163, pp. 322--333. [Online]. Available:
  \url{http://dx.doi.org/10.1007/BFb0034858}
\BIBentrySTDinterwordspacing

\bibitem{Brassard_1987}
\BIBentryALTinterwordspacing
G.~Brassard, C.~Cr{\'e}peau, and J.-M. Robert, ``All-or-nothing disclosure of
  secrets,'' in \emph{Proceedings on Advances in cryptology -- CRYPTO
  '86}.\hskip 1em plus 0.5em minus 0.4em\relax London, UK, UK: Springer-Verlag,
  1987, pp. 234--238. [Online]. Available:
  \url{http://dl.acm.org/citation.cfm?id=36664.36681}
\BIBentrySTDinterwordspacing

\bibitem{Naor_1999_2}
\BIBentryALTinterwordspacing
M.~Naor and B.~Pinkas, ``\BIBforeignlanguage{English}{Oblivious transfer with
  adaptive queries},'' in \emph{\BIBforeignlanguage{English}{Advances in
  Cryptology -- CRYPTO 99}}, ser. Lecture Notes in Computer Science, M.~Wiener,
  Ed.\hskip 1em plus 0.5em minus 0.4em\relax Springer Berlin Heidelberg, 1999,
  vol. 1666, pp. 573--590. [Online]. Available:
  \url{http://dx.doi.org/10.1007/3-540-48405-1\_36}
\BIBentrySTDinterwordspacing

\bibitem{Camenisch_2007}
\BIBentryALTinterwordspacing
J.~Camenisch, G.~Neven, and a.~shelat, ``Simulatable adaptive oblivious
  transfer,'' in \emph{Advances in Cryptology -- EUROCRYPT 2007}, ser. Lecture
  Notes in Computer Science, M.~Naor, Ed.\hskip 1em plus 0.5em minus
  0.4em\relax Springer Berlin Heidelberg, 2007, vol. 4515, pp. 573--590.
  [Online]. Available: \url{http://dx.doi.org/10.1007/978-3-540-72540-4\_33}
\BIBentrySTDinterwordspacing

\bibitem{Green_2008}
\BIBentryALTinterwordspacing
M.~Green and S.~Hohenberger, ``Universally composable adaptive oblivious
  transfer,'' in \emph{Advances in Cryptology - ASIACRYPT 2008}, ser. Lecture
  Notes in Computer Science, J.~Pieprzyk, Ed.\hskip 1em plus 0.5em minus
  0.4em\relax Springer Berlin Heidelberg, 2008, vol. 5350, pp. 179--197.
  [Online]. Available: \url{http://dx.doi.org/10.1007/978-3-540-89255-7\_12}
\BIBentrySTDinterwordspacing

\bibitem{Jarecki_2009}
\BIBentryALTinterwordspacing
S.~Jarecki and X.~Liu, ``\BIBforeignlanguage{English}{Efficient oblivious
  pseudorandom function with applications to adaptive ot and secure computation
  of set intersection},'' in \emph{\BIBforeignlanguage{English}{Theory of
  Cryptography}}, ser. Lecture Notes in Computer Science, O.~Reingold,
  Ed.\hskip 1em plus 0.5em minus 0.4em\relax Springer Berlin Heidelberg, 2009,
  vol. 5444, pp. 577--594. [Online]. Available:
  \url{http://dx.doi.org/10.1007/978-3-642-00457-5\_34}
\BIBentrySTDinterwordspacing

\bibitem{Kurosawa_2009}
\BIBentryALTinterwordspacing
K.~Kurosawa and R.~Nojima, ``\BIBforeignlanguage{English}{Simple adaptive
  oblivious transfer without random oracle},'' in
  \emph{\BIBforeignlanguage{English}{Advances in Cryptology -- ASIACRYPT
  2009}}, ser. Lecture Notes in Computer Science, M.~Matsui, Ed.\hskip 1em plus
  0.5em minus 0.4em\relax Springer Berlin Heidelberg, 2009, vol. 5912, pp.
  334--346. [Online]. Available:
  \url{http://dx.doi.org/10.1007/978-3-642-10366-7\_20}
\BIBentrySTDinterwordspacing

\bibitem{Kurosawa_2010}
\BIBentryALTinterwordspacing
K.~Kurosawa, R.~Nojima, and L.~T. Phong, ``Efficiency-improved fully
  simulatable adaptive ot under the ddh assumption,'' in \emph{Proceedings of
  the 7th International Conference on Security and Cryptography for Networks},
  ser. SCN'10.\hskip 1em plus 0.5em minus 0.4em\relax Berlin, Heidelberg:
  Springer-Verlag, 2010, pp. 172--181. [Online]. Available:
  \url{http://dl.acm.org/citation.cfm?id=1885535.1885554}
\BIBentrySTDinterwordspacing

\bibitem{Green_2011_2}
\BIBentryALTinterwordspacing
M.~Green and S.~Hohenberger, ``\BIBforeignlanguage{English}{Practical adaptive
  oblivious transfer from simple assumptions},'' in
  \emph{\BIBforeignlanguage{English}{Theory of Cryptography}}, ser. Lecture
  Notes in Computer Science, Y.~Ishai, Ed.\hskip 1em plus 0.5em minus
  0.4em\relax Springer Berlin Heidelberg, 2011, vol. 6597, pp. 347--363.
  [Online]. Available: \url{http://dx.doi.org/10.1007/978-3-642-19571-6\_21}
\BIBentrySTDinterwordspacing

\bibitem{Kurosawa_2011}
\BIBentryALTinterwordspacing
K.~Kurosawa, R.~Nojima, and L.~Phong, ``\BIBforeignlanguage{English}{Generic
  fully simulatable adaptive oblivious transfer},'' in
  \emph{\BIBforeignlanguage{English}{Applied Cryptography and Network
  Security}}, ser. Lecture Notes in Computer Science, J.~Lopez and G.~Tsudik,
  Eds.\hskip 1em plus 0.5em minus 0.4em\relax Springer Berlin Heidelberg, 2011,
  vol. 6715, pp. 274--291. [Online]. Available:
  \url{http://dx.doi.org/10.1007/978-3-642-21554-4\_16}
\BIBentrySTDinterwordspacing

\bibitem{Zhang_2013}
\BIBentryALTinterwordspacing
B.~Zhang, H.~Lipmaa, C.~Wang, and K.~Ren,
  ``\BIBforeignlanguage{English}{Practical fully simulatable oblivious transfer
  with sublinear communication},'' in
  \emph{\BIBforeignlanguage{English}{Financial Cryptography and Data
  Security}}, ser. Lecture Notes in Computer Science, A.-R. Sadeghi, Ed.\hskip
  1em plus 0.5em minus 0.4em\relax Springer Berlin Heidelberg, 2013, vol. 7859,
  pp. 78--95. [Online]. Available:
  \url{http://dx.doi.org/10.1007/978-3-642-39884-1\_8}
\BIBentrySTDinterwordspacing

\bibitem{Guleria_2015}
\BIBentryALTinterwordspacing
V.~Guleria and R.~Dutta, ``\BIBforeignlanguage{English}{Efficient adaptive
  oblivious transfer without q-type assumptions in uc framework},'' in
  \emph{\BIBforeignlanguage{English}{Information and Communications Security}},
  ser. Lecture Notes in Computer Science, L.~C.~K. Hui, S.~H. Qing, E.~Shi, and
  S.~M. Yiu, Eds.\hskip 1em plus 0.5em minus 0.4em\relax Springer International
  Publishing, 2015, vol. 8958, pp. 105--119. [Online]. Available:
  \url{http://dx.doi.org/10.1007/978-3-319-21966-0\_8}
\BIBentrySTDinterwordspacing

\bibitem{Crepeau_2005}
\BIBentryALTinterwordspacing
C.~Cr{\'e}peau, K.~Morozov, and S.~Wolf,
  ``\BIBforeignlanguage{English}{Efficient unconditional oblivious transfer
  from almost any noisy channel},'' in
  \emph{\BIBforeignlanguage{English}{Security in Communication Networks}}, ser.
  Lecture Notes in Computer Science, C.~Blundo and S.~Cimato, Eds.\hskip 1em
  plus 0.5em minus 0.4em\relax Springer Berlin Heidelberg, 2005, vol. 3352, pp.
  47--59. [Online]. Available:
  \url{http://dx.doi.org/10.1007/978-3-540-30598-9\_4}
\BIBentrySTDinterwordspacing

\bibitem{Rivest_1999}
R.~Rivest, ``Unconditionally secure commitment and oblivious transfer schemes
  using private channels and a trusted initializer,'' Unpublished manuscript,
  1999.

\bibitem{Yao_1982_2}
A.~C. Yao, A.~C. Yao, A.~C. Yao, and A.~C. Yao, ``Protocols for secure
  computations,'' in \emph{Foundations of Computer Science, 1982. SFCS '08.
  23rd Annual Symposium on}, Nov 1982, pp. 160--164.

\bibitem{Goldreich_1987}
\BIBentryALTinterwordspacing
O.~Goldreich, S.~Micali, and A.~Wigderson, ``How to play any mental game,'' in
  \emph{Proceedings of the Nineteenth Annual ACM Symposium on Theory of
  Computing}, ser. STOC '87.\hskip 1em plus 0.5em minus 0.4em\relax New York,
  NY, USA: ACM, 1987, pp. 218--229. [Online]. Available:
  \url{http://doi.acm.org/10.1145/28395.28420}
\BIBentrySTDinterwordspacing

\bibitem{Shannon_1949}
\BIBentryALTinterwordspacing
C.~E. Shannon, ``Communication theory of secrecy systems*,'' \emph{Bell System
  Technical Journal}, vol.~28, no.~4, pp. 656--715, 1949. [Online]. Available:
  \url{http://dx.doi.org/10.1002/j.1538-7305.1949.tb00928.x}
\BIBentrySTDinterwordspacing

\bibitem{Katz_2007}
J.~Katz and Y.~Lindell, \emph{Introduction to Modern Cryptography}.\hskip 1em
  plus 0.5em minus 0.4em\relax Chapman \& Hall/CRC, 2007.

\end{thebibliography}

\end{document}